%% file: main.tex
\documentclass[submission,copyright,creativecommons]{eptcs}
\providecommand{\event}{FOCLASA 2014} 
\usepackage{breakurl}             

\usepackage{csquotes}
\usepackage{graphicx}
\usepackage{balance}
\usepackage{AMMALanguages}
\usepackage{longtable}
\usepackage{url}
\usepackage{lscape}
\usepackage[named]{algo}

\usepackage{xcolor}
\usepackage{amsmath}
\usepackage{amssymb}
\usepackage{amsxtra}

\usepackage{colortbl,multirow,hhline}

\usepackage{ifthen}
\newboolean{showcomments}
\setboolean{showcomments}{true} 
\ifthenelse{\boolean{showcomments}}
  {\newcommand{\nb}[2]{
    \fcolorbox{gray}{yellow}{\bfseries\sffamily\scriptsize#1}
    {\sf\small$\blacktriangleright$\textit{#2}$\blacktriangleleft$}
   }
   
  }
  {\newcommand{\nb}[2]{}
   
  }

\newtheorem{definition}{Definition}
\newcommand{\freccia}[1]{\mathop{\stackrel{#1} {\longrightarrow}} }
\newcommand{\echain}[1]{\mathop{\stackrel{#1} {\Longrightarrow}} }

\urldef{\mailsa}\path|{marco.autili,massimo.tivoli}@univaq.it|

\title{Distributed Enforcement of Service Choreographies}

\author{Marco Autili \qquad\qquad Massimo Tivoli
\institute{Department of Information Engineering Computer Science and Mathematics}
\institute{University of L'Aquila - ITALY}
\email{marco.autili@univaq.it \quad\qquad massimo.tivoli@univaq.it}
}

\def\titlerunning{Distributed Enforcement of Service Choreographies}
\def\authorrunning{M. Autili \& M. Tivoli}

\begin{document}
\maketitle

\begin{abstract}
Modern service-oriented systems are often built by reusing, and composing together, existing services distributed over the Internet. Service choreography is a possible form of service composition whose goal is to specify the interactions among participant services from a global perspective. In this paper, we formalize a method for the distributed and automated enforcement of service choreographies, and prove its correctness with respect to the realization of the specified choreography. The formalized method is implemented as part of a model-based tool chain released to support the development of choreography-based systems within the EU CHOReOS project. We illustrate our method at work on a distributed social proximity network scenario.
\end{abstract}

\input{introduction}
\input{scenario}
\input{preamble}
\input{enforcement}

\input{correctness}
\input{related}
\input{conclusions}

\bibliographystyle{eptcs}
\bibliography{main,biblio2,biblio3}
\end{document}

%% file: introduction.tex
\section{Introduction} \label{sec:introduction}

\vspace{-0.3cm}

The future trend in service-oriented development is to build systems by reusing, and composing together, existing services distributed over the Internet.

Service choreography is a form of service composition whose goal is to specify message exchanges among multiple partner services, called {\em participants}, from a global perspective. When building a service-based system, a possible engineering approach
is to compose services in a distributed way by considering this global specification. To this extent, the following two problems are usually considered: (a) {\em realizability check} -- checks whether the choreography can be realized by implementing each participant so that it conforms to the choreography role that specifies its ``expected'' behaviour; and (b) {\em conformance check} -- checks whether the global interaction of a set of services satisfies the choreography. In the literature many approaches have been proposed to address these problems, e.g.,~\cite{CGL08,Sal08,pascal12,Bultan:2011,Basu-Bultan-POPL:12}. However, when the goal is to actually realize a service choreography by reusing {\em third-party} services, hence going beyond just checking its effectiveness, a further problem worth to be considered concerns automated choreography {\em enforcement}. That is, how to coordinate the interactions among participant services in order to fit the choreography specification. This requires to distribute and enforce, among the participants, the global coordination logic extracted from the choreography specification. The general problem here is that, although services may have been discovered or registered as suitable participants, their composite interaction may prevent the choreography realization if left uncontrolled (or wrongly coordinated).

BPMN2\footnote{\url{http://www.omg.org/spec/BPMN/2.0}} {\em Choreography Diagrams} represent a de facto standard in the current practice of choreography specification and design. Within the CHOReOS EU project\footnote{\url{http://www.choreos.eu}}, the work presented in this paper is part of, we implemented a model-based transformation\footnote{See documentation available at \url{choreos.disim.univaq.it} for details.} to synthesize, out of a BPMN2 choreography diagram, an intermediate state-based model called {\em Choreography explicit-Flow Model} (CeFM). The latter is a choreography model that, conforming to the BPMN2 standard specification, makes explicit coordination-related information that in BPMN2 is implicit. This allows to statically infer the information needed for enabling distributed coordination that, otherwise, should be calculated at run time for each choreography instance and for each execution of it. For instance, the CeFM model specifies the source and target state from which a task is initiated and terminated, the corresponding transition and enabling condition.


\vspace{-0.3cm}

\begin{figure}[h]
  \center
    \includegraphics[width=\textwidth]{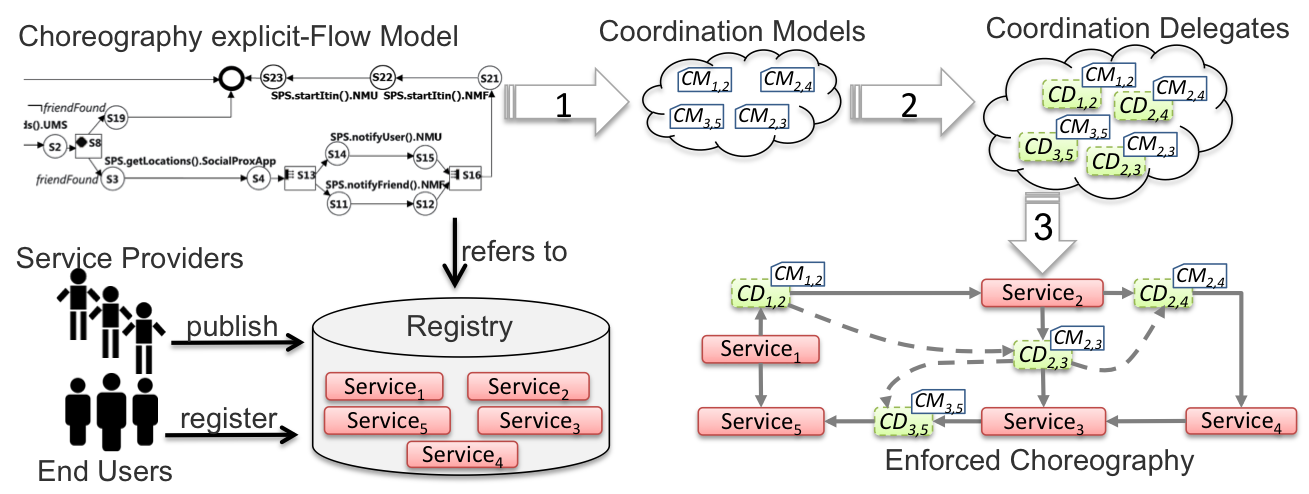}
    \vspace{-0.9cm}
\caption{Overview of the choreography enforcement method}
  \label{fig:ArchSample}
\end{figure}


\noindent {\bf Contribution of the paper} -- In this paper, we formalize a choreography enforcement method based on distributed coordination of the participants from outside. Figure~\ref{fig:ArchSample} shows an overview of our method as organized into three steps. Out of a CeFM $C$, our method automatically derives a set of what we call {\em Coordination Delegates} (CDs). In particular, the coordination logic modeled by $C$ is distributed into a set of {\em Coordination Models} (CMs), one for each CD (step 1 in the figure). The CDs exploit the CMs in order to coordinate the interaction of the participants in a way that the resulting collaboration realizes the choreography specified by $C$ (step 2 in the figure). The derived CDs are interposed (only when strictly needed) among the participants according to the CHOReOS architectural style~\cite{fase:2013,D1.4b} (step 3 in the figure). To give an intuition, Figure~\ref{fig:ArchSample} uses a box/arrow (i.e., component/connector) notation to show a sample architecture instance conforming to the CHOReOS style (see the bottom-most diagram in the right-hand side of the figure). Within CHOReOS, our method refers to a service registry to discover services that are suitable to play the roles of the choreography. The registry contains services published by providers that have identified business opportunities in the domain of interest. The registry also contains the registration of the end users interested in exploiting the choreography through client applications.

CDs perform pure business-level coordination by proxifying the service interaction, 
and mediate it by exchanging the coordination information 
contained in their own CMs, in a fully distributed way. In this way, CDs prevent possible {\em undesired interactions}. The latter are those interactions that do not belong to the set of interactions allowed by $C$ and can happen when the services collaborate in an uncontrolled way. We formalize the notions of CeFM and CM, and show how to decompose a CeFM into a set of CMs. Furthermore, we describe the distributed coordination algorithm that is implemented by the CDs to realize the coordination logic. 

The overall enforcement approach has been implemented as a set of REST services, and a graphical Eclipse plugin that invokes them has been released\footnote{\url{choreos.disim.univaq.it} and \url{www.ow2.org/view/Future_Internet}}. A tool demo screencast is also available.

\newpage

\noindent {\bf Advance with respect to our previous work} -- The work described in this paper represents an advance with respect to our previous work in~\cite{fase:2013}. In fact, although the synthesis process described there treats most of the BPMN2 constructs, it considers a simplified version of their actual semantics. For instance, the selection of conditional branches is simply abstracted as a non-deterministic choice, regardless the run-time evaluation of their enabling conditions. Analogously, parallel flows are enforced by non-deterministically choosing one of their linearizations obtained through interleaving, hence loosing the actual parallelism degree. 
To overcome these limitations, in this paper, we formalize the CeFM model that, being more expressive than the choreography model adopted in~\cite{fase:2013}, preserves the actual semantics of the BPMN2 constructs. Relying on the CeFM model has led us to define a novel and more effective distributed coordination algorithm. As a further advance with respect to~\cite{fase:2013}, we also prove correctness of the algorithm with respect to preventing undesired interactions. In particular, the proof gives a rigorous characterization of the notion of undesired interaction that in~\cite{fase:2013} has been informally treated hence making it difficult to effectively assess our method.
%
%
%
%
%
%

\smallskip
\noindent {\bf Main focus of the paper} -- It is worth to mention that the coordination logic performed by the CDs is {\em service-independent} since it is based on the expected behaviour of the participants as specified by $C$, rather than on the actual one as obtained after discovery. Within CHOReOS, this is done to consistently realize {\em separation of concerns}. That is, to separate pure coordination issues (i.e., undesired interactions) from adaptation ones (e.g., syntactic mismatches at the service interface level). The latter can arise whenever a service discovered as a participant does not exactly match the role to be played. Adaptation issues, as well as discovery ones, are out of scope for this paper and we refer to~\cite{serene:2013,icse:2013} for details about their treatment within CHOReOS.

%
%

\smallskip
\noindent {\bf Structure of the paper} -- Section~\ref{sec:scenario} introduces a {\em distributed social proximity network} scenario that is used as illustrative example. Section~\ref{sec:preamble} formalizes the notion of CeFM. Its decomposition into a set of CMs is described in Section~\ref{sec:preamble-bis}. Section~\ref{sec:enforcement} formalizes a distributed coordination algorithm, which describes the coordination logic that a CD has to perform by relying on its CM. We prove correctness of the distributed coordination algorithm in Section~\ref{sec:correctness}. To this extent, in Section~\ref{sec:correctness}, we also formalize the notion of undesired interactions. Section~\ref{sec:related} discusses related works. Conclusions and final remarks are given in Section~\ref{sec:conclusions}. 

%% file: scenario.tex
\section{The distributed social proximity network scenario} \label{sec:scenario}



\vspace{-0.3cm}

\begin{figure}[h]
  \center
    \includegraphics[width=1.0\textwidth]{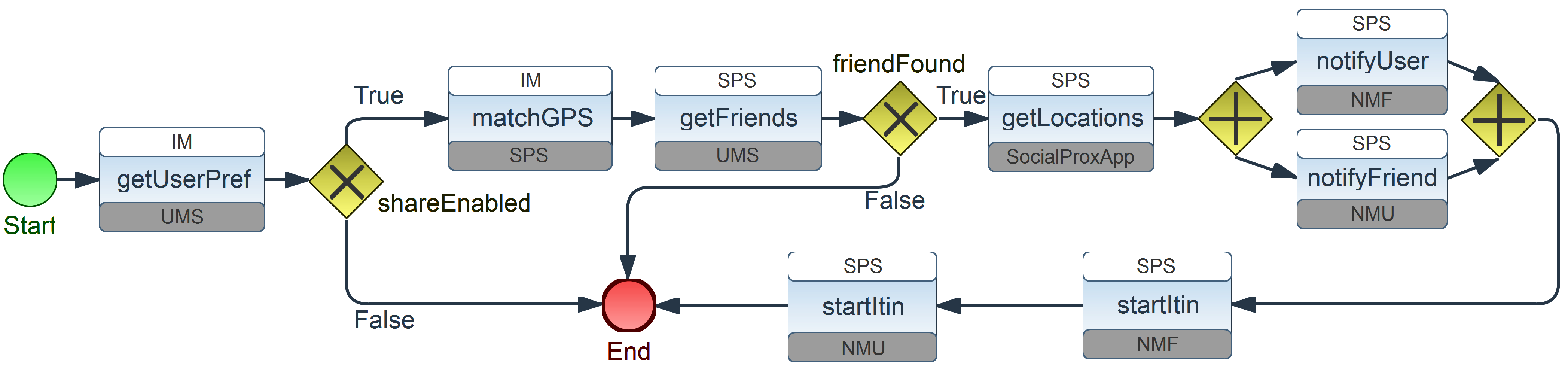}
    \vspace{-0.8cm}
\caption{BPMN2 choreography diagram of the Distributed Social Proximity Network scenario}
  \label{fig:bpmn2Sample}
\end{figure}

With reference to Figure~\ref{fig:bpmn2Sample}, choreography diagrams use rounded-corner boxes to
denote choreography {\em tasks} (e.g., {\tt getUserPref}). Each of them is labeled with the roles of the two participants
involved in the task, and the name of the {\em service operation} performed by the initiating
participant and provided by the other one. A role contained in a white box
denotes the initiating participant (e.g., {\tt IM}); a role contained in a filled box denotes the receiving participant (e.g., {\tt UMS}).


\noindent The choreography in the figure models a small portion of a  \emph{Distributed Social Proximity Network} scenario\footnote{Within CHOReOS, the scenario is part of a Dynamic Route system, which has been the pilot use case that we used to validate the overall choreography synthesis process.}, which concerns the collaboration of location-based services and citizen mobile apps. The choreography considers user preferences, user friend lists, and user location under predefined private data policy, to create ad-hoc itineraries between two citizens that know each other and are willing to meet in some place.
In brief, upon receiving a start event from a citizen (i.e., from the dedicated mobile app named \texttt{SocialProxApp}), the Itinerary Manager (\texttt{IM}) calls the User Manager Service (\texttt{UMS}) to check if location sharing is enabled according to the citizen preferences. If yes (see the conditional branching represented as a rhombus marked with a ``x''), \texttt{IM} starts matching the GPS position of all citizen contacts by using the Social Proximity Service (\texttt{SPS}). Then, after interacting with \texttt{UMS} to get the list of all friends found nearby with location sharing enabled, \texttt{SPS} interacts with the \texttt{SocialProxApp}s of the friends in the list to get their location. Finally, after a friend has been selected by the requesting citizen among the displayed list of available contacts, and the friend has accepted, \texttt{SPS} notifies in parallel the Notification Managers, \texttt{NMU} and \texttt{NMF}, for both the requesting user and the selected friend, respectively (see the parallel branch represented as an rhombus marked with a ``+'' with two outgoing arrows). These parallel flows are joined afterwards as soon as both the notification tasks have been accomplished (see the merging branch represented as an rhombus marked with a ``+'' with two incoming arrows), hence letting the two friends start their itineraries. An important aspect of this simple scenario is that the choreography in Figure~\ref{fig:bpmn2Sample} specifies that \texttt{SPS} must notify both the user and the friend before allowing them to start their respective itineraries. The choreography uses the merging branch at the right-hand side of the figure to model this desired behavior. When enforcing this choreography specification by using existing third-party services as participants, possible undesired interactions violating the specification can occur. For instance, the service playing the role of \texttt{SPS} attempts at notifying the user and starting his itinerary before notifying the friend.


\vspace{-0.3cm}

\begin{figure}[h]
  \center
    \includegraphics[width=1.0\textwidth]{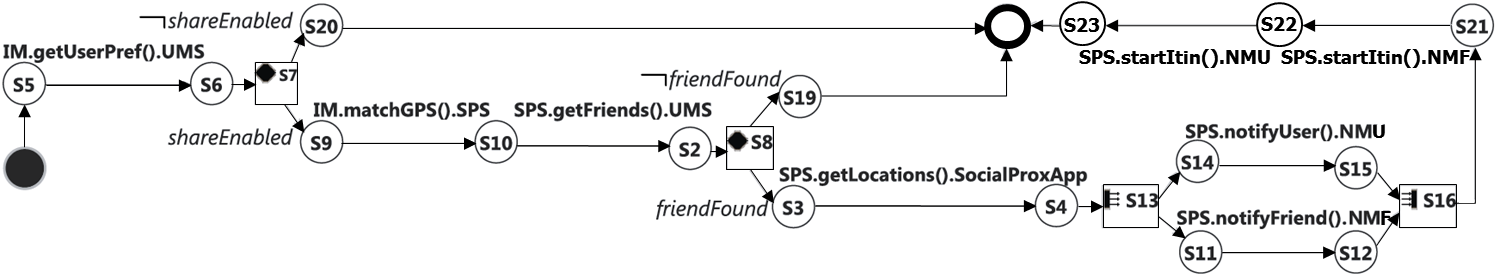}
    \vspace{-0.8cm}
\caption{CeFM derived from the BPMN2 choreography diagram in Figure~\ref{fig:bpmn2Sample}}
  \label{fig:cltsSample}
\end{figure}

\vspace{-0.3cm}

\noindent Figure~\ref{fig:cltsSample} shows the CeFM automatically derived out of the diagram in Figure~\ref{fig:bpmn2Sample}. The numbering of the states in the CeFM depends on the performed model-based transformation. The symbol ``$\neg$'' denotes the classical negation in propositional logic. Transitions are possibly labeled by the initiating participant, the performed task, and the receiving participant, separated by `{\tt .}'.

In the following two sections, we formalize the notion of CeFM and its decomposition into a set of CMs, respectively.


%% file: preamble.tex
\section{Choreography explicit-Flow Model} \label{sec:preamble}


\vspace{-0.3cm}

Let $\mathcal{T}$ be the universal set of choreography tasks (e.g., service operations). 
Let $\mathcal{R}$ be the universal set of choreography roles. 
Let $\Phi$ be the universal set of Propositional Logic (PL) formulae. PL formulae are used to distinguish choreography flows depending on whether a specific condition, represented as a {\em predicate} in PL, holds or not. More technically, according to the BPMN2 standard specification, in our implementation, a predicate is a conditional expression given in the syntax of the Java language. Thus, all data values involved in a predicate evaluation are assumed to actually range over finite domains.

The following definition characterizes the structural properties of a CeFM as required for enabling automated synthesis of CDs. As already introduced in Section~\ref{sec:introduction}, the intuition behind Definition~\ref{clts_def} is that a BPMN2 choreography diagram has to be transformed into an equivalent choreography model that, within the CDs synthesis process, allows to statically infer coordination-related information.

\begin{definition}[Choreography explicit-Flow Model] \label{clts_def} A {\it Choreography explicit-Flow Model} (CeFM) $C$ is a tuple $($$S^{ALT}_C$$,$$S^{LOOP}_C$$,$$S^{FORK}_C$$,$$S^{JOIN}_C$$,$$S^{PLAIN}_C$$,$$s^0_C$$,$$s^f_C$$,$$R_C$$,$$T_C$$,$$D_C$$)$, where $S^{ALT}_C$$,$$S^{LOOP}_C$$,$$S^{FORK}_C$$,$$S^{JOIN}_C$ and $S^{PLAIN}_C$ are disjoint and finite sets of states denoting {\em alternative}, {\em loop}, {\em fork}, {\em join}, and {\em plain} states, respectively. $s^0_C$ and $s^f_C$ denote the {\em initial} (i.e., with no incoming transitions) and {\em final} (i.e., with no outgoing transitions) states, respectively, and they are neither alternative, loop, fork, join, nor plain states. Let $S_C$ be the union set of all kinds of state, i.e., $S_C$$=$$S^{ALT}_C$$\cup$$S^{LOOP}_C$$\cup$$S^{FORK}_C$$\cup$$S^{JOIN}_C$$\cup$$S^{PLAIN}_C$$\cup$$\{s^0_C,s^f_C\}$. $R_C$ is a finite set of roles, i.e., $R_C$$\subseteq$$\mathcal{R}$. 
$T_C$$\subseteq$$($$\mathcal{R}$ 
$\times$$\mathcal{T}$ 
$\times$$\mathcal{R}$ 
$)$$\cup$$\Phi$ is a set of task labels called the {\it alphabet} of $C$. $D_C \subseteq S_C \times T_C \cup \{\varepsilon\} \times S_C$ is the flow relation. $\varepsilon$ denotes an $\varepsilon$-task (i.e., an ``empty'' task). $C$ is finite if $D_C$ is finite and $C$ is empty if $D_C$ is empty. We will make use of the following notations: (i) $g \freccia{\alpha}_C h$ iff $(g,\alpha,h)\in D_C$ $\wedge$ $\alpha$=$r.t.r^\prime$ for some $r,r^\prime$$\in$$R_C$ 
, $t\in \mathcal{T}$; (ii) $g \freccia{\rho}_C h$ iff $(g,\rho,h)\in D_C$ for some $\rho$$\in$$\Phi$; and (iii) $g \freccia{}_C h$ iff $(g,\varepsilon,h)\in D_C$.  
In all the cases we will write that $g$ is a {\em predecessor} of $h$ or, equivalently, $h$ is a {\em successor} of $g$. The following structural properties hold:
\begin{itemize}
\item {\bf plain}: {\it for all} $s^{plain}$$\in$$S^{PLAIN}_C$, $s^{plain}$ {\it has only one predecessor} $s$$\in$$S_C$$\setminus$$\{s^f_C\}$ {\it such that} $s$$\freccia{}_C$$s^{plain}$ {\it and} $s^{plain}$ {\it has only one successor} $s^\prime$$\in$$S_C$$\setminus$$\{s^0_C\}$ {\it such that} $s^{plain}$$\freccia{}_C$$s^\prime$;

\item {\bf alternative:} {\it for all} $s^{alt}$$\in$$S^{ALT}_C$, $s^{alt}$ {\it has only one predecessor} $s$$\in$$S_C$$\setminus$$\{s^f_C\}$ {\it such that} $s$$\freccia{}_C$$s^{alt}$ {\it and} $s^{alt}$ {\it has at least two successors} $s_1$,$s_2$$\in$$S_C$$\setminus$$\{s^0_C\}$ {\it such that} $s^{alt}$$\freccia{\rho_1}_C$$s_1$ and $s^{alt}$$\freccia{\rho_2}_C$$s_2$ {\it for some} $\rho_1$,$\rho_2$$\in$$\Phi$; {\it for all pairs of successors} $(s^\prime_i,s^\prime_j)$, {\it with} $i\neq j$, {\it and respective predicates} $\rho_i$,$\rho_j$$\in$$\Phi$ {\it such that} $s^{alt}$$\freccia{\rho_i}_C$$s^\prime_i$ {\it and} $s^{alt}$$\freccia{\rho_j}_C$$s^\prime_j$ {\it then there is no variable assignment that makes} $\rho_i$$\wedge$$\rho_j$ hold;
\item {\bf loop:} {\it for all} $s^{loop}$$\in$$S^{LOOP}_C$, $s^{loop}$ {\it has only two predecessors} $s,s^\prime$$\in$$S_C$$\setminus$$\{s^f_C\}$ {\it such that} $s$$\freccia{}_C$$s^{loop}$ {\it and} $s^\prime$$\freccia{}_C$$s^{loop}$, {\it and} $s^{loop}$ {\it has only two successors} $q,q^\prime$$\in$$S_C$$\setminus$$\{s^0_C\}$ {\it such that} $s^{loop}$$\freccia{}_C$$q$, $s^{loop}$$\freccia{\rho}_C$$q^\prime$ {\it for some} $\rho$$\in$$\Phi$, {\it and} $s^\prime$ is reachable from $q^\prime$; {\it we call} $q^\prime$ {\it and} $q$ {\it the} {\em loop entry state} {\it and} {\em exit state}, {\it respectively};
\item {\bf fork:} {\it for all} $s^{fork}$$\in$$S^{FORK}_C$, $s^{fork}$ {\it has only one predecessor} $s$$\in$$S_C$$\setminus$$\{s^f_C\}$ {\it such that} $s$$\freccia{}_C$$s^{fork}$ {\it and} $s^{fork}$ {\it has at least two successors} $s^\prime$,$s^{\prime\prime}$$\in$$S_C$$\setminus$$\{s^0_C\}$ {\it such that} $s^{fork}$$\freccia{}_C$$s^\prime$ {\it and} $s^{fork}$$\freccia{}_C$$s^{\prime\prime}$;
\item {\bf join:} {\it for all} $s^{join}$$\in$$S^{JOIN}_C$, $s^{join}$ {\it has only one successor} $s$$\in$$S_C$$\setminus$$\{s^0_C\}$ {\it such that} $s^{join}$$\freccia{}_C$$s$ {\it and} $s^{join}$ {\it has at least two predecessors} $s^\prime$,$s^{\prime\prime}$$\in$$S_C$$\setminus$$\{s^f_C\}$ {\it such that} $s^\prime$$\freccia{}_C$$s^{join}$ {\it and} $s^{\prime\prime}$$\freccia{}_C$$s^{join}$.
\end{itemize}
\end{definition}

\noindent BPMN2 specification employs the theoretical concept of {\em token} that, traversing the sequence flows and passing through the elements in a BPMN2 process, aids to define its behavior. In line with this concept, the formal semantics of a CeFM is given in Section~\ref{sec:preamble-bis} as a set of coordination models (see Definition~\ref{cmg_def}) obtained through the automatic generation process formalized by Definition~\ref{cmgen_def}.
In particular, the generation process leverages the notion of \emph{structural sequential flow} in Definition~\ref{flow_def} to formally model all the paths traversing the choreography diagram. The coordination models are meant to coordinate these paths according to the different state types.
%
%

Intuitively, the initial state of a CeFM generates the token that must eventually be consumed at the final state. If a token is on a plain state, then the outgoing flow is ready to be executed hence performing the operation characterized by its task label. If a token is on an alternative state, then one of the outgoing flows whose predicate holds is ready to be executed. This means that, according to the above structural properties, we deal with BPMN2 Diverging (Decision) Exclusive Gateways through alternative states whose outgoing flows are guarded by mutually exclusive predicates, i.e., at run time only one predicate evaluates to true\footnote{Indeed, in our implementation we also allow the specification of default a branch, not considered here for simplicity.}. At the same time, according to the BPMN2 standard specification, we disallow the specification of non deterministic gateways with no condition specified.
If a token is on a loop state, then the outgoing flow leading to the loop entry state is ready to be executed if its guarding predicate holds. Otherwise, the flow leading to the exit state will be performed. If a token is on a fork state, then all outgoing flows are taken simultaneously leading to several parallel executions, each having its own token. If, for each of its incoming flows, there is a token on a join state, then the outgoing flow is triggered.

Due to the introduction of 
$\varepsilon$-tasks, in general, a CeFM is non-deterministic. $\varepsilon$-tasks are introduced for dealing with non-plain states both structurally and semantically. For instance, they can be used to specify cascading fork states or to model the (internal) event of creating parallel flows, which originate from a fork state.

By leveraging the concept of token, alternative models, such as free-choice Petri nets~\cite{fcPNs}, might have been adopted. However, the deep study we have initially conducted within CHOReOS to precisely define, at the project level, the integration architecture for the CHOReOS Integrated Development and Run-time Environment\footnote{ \url{http://www.choreos.eu/bin/view/Documentation/WebHome}} (IDRE), led the whole consortium to agree on the definition of the CeFM model, which best met the (both formal and technical) requirements of all the software tools now integrated by the IDRE. Indeed, to the purposes of defining an integrated suite of tools to support the whole choreography life cycle, the CeFM model brings together many features of already existing formalisms and notations in the literature, and filters out those ones not strictly needed.
Last but not least, one of the main requirements was to have a notation as close as possible to the BPMN2 choreography diagrams, while enabling formal reasoning and fully automatic treatment by all the IDRE components.

\section{Automated decomposition of a CeFM into a set of CMs} \label{sec:preamble-bis}

\vspace{-0.3cm}

From hereon let $C$$=$$($$S^{ALT}_C$$,$$S^{LOOP}_C$$,$$S^{FORK}_C$$,$$S^{JOIN}_C$$,$$S^{PLAIN}_C$$,$$s^0_C$$,$$s^f_C$$,$$R_C$$,$$T_C$$,$$D_C$$)$ be a CeFM. The ability of explicitly identifying specific structural sequential flows of $C$ represents a basic ingredient to support the automated synthesis of the CDs that cooperatively realize $C$. For instance, in order to deal with the synchronization of several parallel flows in correspondence of a join state, it is sufficient for a CD to extract out of $C$ each of these flows as originating from the same fork. This allows a CD, reaching a join state, to be aware of (i) those CDs that must be notified about the reach of the join state and, complementary, (ii) those states (predecessors of the join state) that must be waited for.

\begin{definition}[Structural sequential flow] \label{flow_def} The sequence $t$$=$$\alpha_{i+1}$$\alpha_{i+2}$$\ldots$$\alpha_n$ is a {\em structural sequential flow} of $C$ iff there exists a sequence of states $s_i$$\ldots$$s_n$$\in$$S_C$ such that $s_i$$\freccia{\alpha_{i+1}}_C$$s_{i+1}$$\ldots$$s_{n-1}$$\freccia{\alpha_n}$$s_n$, $i\geq 0$, $n > i$. The {\em empty flow} is denoted by $\dddot{\varepsilon}$. We will make use of the following notation: $s_1 \echain{\alpha}_C s_n$ iff {\it there exists} $n\geq 2$ {\it such that} $s_1$$\freccia{\alpha}_C$$s_2$$\freccia{}_C$$\ldots$$\freccia{}_C$$s_n$.
\end{definition}

\noindent As introduced in Section~\ref{sec:introduction}, a further basic ingredient of our synthesis method concerns the distribution step for $C$. That is, $C$ is distributed into a set of CMs, one for each CD. We recall that CDs are not always generated for each pair of participants. Indeed, for two given participants, a CD is generated only when there is a dependency relation between them. This means that we do not require to always keep the global state of the choreography to enforce it. This characterizes the distributed nature of our synthesis approach.

Each CM represents a local view of $C$. It is local since it models the coordination logic that has to be enforced on the interaction between two participants, $p_i$ and $p_j$. This is done by exchanging coordination information among the CD supervising $p_i$ and the CDs it needs to synchronize with. Thus, the set of all CMs can be considered as a distributed model of $C$. 
A CM characterizes the coordination information that is needed to automatically synthesize the coordination logic that the corresponding CD has to perform, while interacting with the other CDs, in order to realize $C$.

\begin{definition}[Coordination Model] \label{cmg_def}
Let $i,j$$\in$$\mathbb{N}$, with $i\neq j$, be the identifiers of two participant services. Then, the {\em Coordination Model} $CM_{i,j}$ for the operations required by the participant $i$ and provided by the participant $j$ is the set\footnote{As usual, let $S$ be a set, $2^S$ denotes the power-set of $S$} of tuples $\{ \tau \mid \tau\in S_C \times T_C \times S_C \times 2^{\mathbb{N}\times \mathbb{N}} \times \Phi \times 2^{S_C\times \mathbb{N}\times \mathbb{N}} \times 2^{S_C\times \mathbb{N}\times \mathbb{N}}\}$. For each tuple $\langle$$s$$,$$t$$,$$s^\prime$$,$$CD_{s^\prime}$$,$$\rho$$,$$Notify_s$$,$$Wait_{siblings(s)}$$\rangle$ {\em :}

\begin{itemize}
\item $s$ denotes the CeFM source state from which the related CD can either perform the operation $t$ or take a move without performing any operation (i.e., the CD can step over an $\varepsilon$-task). In both cases, $s^\prime$ denotes the reached target state;

\item $CD_{s^\prime}$ contains the set of (identifiers of) those CDs whose supervised services became active in $s^\prime$, i.e., the ones that will be allowed to require/provide some operation from $s^\prime$. This information is used by the ``currently active'' CDs to inform the set of ``to be activated'' CDs (in the target state) about the changing global state;

\item $\rho$ is a PL predicate whose validity has to be checked to select the correct tuple, and hence the correct flow(s) in the CeFM;

\item $Notify_s$ contains the predecessor of a join state that a CD, when reaching it, must notify to the other CDs in the parallel flow(s) of the same originating fork. Complementary, $Wait_{siblings(s)}$ contains the predecessors of join states that must be waited for.
\end{itemize}
\end{definition}

\noindent According to the above definition, each $CM_{i,j}$ is automatically synthesized out of $C$ as formalized by the following definition.

\begin{definition}[Coordination Model Generation] \label{cmgen_def}
Let $1,\ldots,n$ be identifiers for $n$ participant services playing the roles $r_1,\ldots,r_n$$\in$$\mathcal{R}$, respectively. Then, $CM_{i,j}$ {\em is generated out of} $C$ as follows:

\indent\indent$CM_{i,j}$$=$$\{$$\langle$$s$$,$$t$$,$$s^\prime$$,$$CD_{s^\prime}$$,$$true$$,$$Notify_s$$,$$Wait_{siblings(s)}$$\rangle$$|$ $s$$\freccia{r_i.t.r_j}_C$$s^\prime$$\}$ $\bigcup$

\indent\indent\indent\indent $\{$$\langle$$s$$,$$\varepsilon$$,$$s^\prime$$,$$CD_{s^\prime}$$,$$true$$,$$Notify_s$$,$$Wait_{siblings(s)}$$\rangle$$|$$\exists$$s^{prec}, t$$:$ $s^{prec}$$\echain{r_i.t.r_j}_C$$s$$\freccia{}_C$$s^\prime$$\}$ $\bigcup$

\indent\indent\indent\indent $\{$$\langle$$s$$,$$\varepsilon$$,$$s^\prime$$,$$CD_{s^\prime}$$,$$\rho_{s.s^\prime}$$,$$Notify_s$$,$$Wait_{siblings(s)}$$\rangle$$|$
$\exists$$s^{prec}, t$$:$ $s^{prec}$$\echain{r_i.t.r_j}_C$$s$$\freccia{\rho_{s.s^\prime}}_C$$s^\prime$$\}$

\noindent where

\indent\indent $CD_{s^\prime}$$=$ $\{$$(h,k)$$|$$(h,k)$$\neq$$(i,j)$$\wedge$$\exists s^{\prime\prime},t$$:$ $s^\prime$$\freccia{r_h.t.r_k}_C$$s^{\prime\prime}$ $\}$ $\bigcup$

\indent\indent\indent\indent $\{$$(h,k)$$|$$(h,k)$$\neq$$(i,j)$$\wedge$$\exists \rho$$\in$$\Phi, s^{\prime\prime}, s^{succ}, t$$:$ $s^\prime$$\freccia{\rho}_C$$s^{\prime\prime}$$\freccia{r_h.t.r_k}_C$$s^{succ}$ $\}$ $\bigcup$

\indent\indent\indent\indent $\{$$(h,k)$$|$$(h,k)$$\neq$$(i,j)$$\wedge$$s^\prime$$\in$$S^{FORK}_C$$\cup$$S^{JOIN}_C$$\wedge$$\exists s^{\prime\prime}, s^{succ},t$$:$ $s^\prime$$\freccia{}_C$$s^{\prime\prime}$$\freccia{r_h.t.r_k}$$s^{succ}$ $\}$

\indent\indent $Notify_s$$=$$\{$$[s,(h,k)]$$|$$s^\prime$$\in$$S^{JOIN}_C$$\wedge$$\exists s^{\prime\prime}$$\neq$$s, s^{prec}, t:$$s^{prec}$$\echain{r_h.t.r_k}_C$$s^{\prime\prime}$$\freccia{}_C s^\prime \}$

\indent\indent $Wait_{siblings(s)}$$=$ $\{$$[s^{\prime\prime},(h,k)]$$|$$s^\prime$$\in$$S^{JOIN}_C$$\wedge$$\exists s^{\prime\prime}$$\neq$$s, s^{prec}, t:$$s^{prec}$$\echain{r_h.t.r_k}_C$$s^{\prime\prime}$$\freccia{}_C s^\prime \}$
\end{definition}


\begin{table}[h]
\begin{center}
{
\scriptsize
\begin{tabular}{|l|l|}
\hline

         \cellcolor{gray!25}$\textbf{CM}_{\textbf{IM,UMS}}$ & \cellcolor{gray!25}$\textbf{CM}_{\textbf{SPS,UMS}}$ \\

\hline

        \ \ \ \ $\langle S5, getUserPref(), S6, \{\}, true, \{\}, \{\} \rangle$ & \ \ \ \ $\langle S10, getFriends(), S2, \{\}, true, \{\}, \{\} \rangle$ \\

        \ \ \ \ $\langle S6, \varepsilon, S7, \{\}, true, \{\}, \{\} \rangle$ &
        \ \ \ \ $\langle S2, \varepsilon, S8, \{\}, true, \{\}, \{\} \rangle$ \\

        \ \ \ \ $\langle S7, \varepsilon, S20, \{\}, \neg shareEnabled, \{\}, \{\} \rangle$ &
        \ \ \ \ $\langle S8, \varepsilon, S19, \{\}, \neg friendFound, \{\}, \{\} \rangle$ \\

        \ \ \ \ $\langle S7, \varepsilon, S9, \{(IM,SPS)\}, shareEnabled, \{\}, \{\} \rangle$ &
        \ \ \ \ $\langle S8, \varepsilon, S3, \{(SPS,SocialProxApp)\}, friendFound, \{\}, \{\} \rangle$ \\

        \ \ \ \ $\langle S20, \varepsilon, Final, \{\}, true, \{\}, \{\} \rangle$ &
        \ \ \ \ $\langle S19, \varepsilon, Final, \{\}, true, \{\}, \{\} \rangle$ \\

\hline
         \cellcolor{gray!25}$\textbf{CM}_{\textbf{IM,SPS}}$ &
         \cellcolor{gray!25}$\textbf{CM}_{\textbf{SPS,SocialProxApp}}$ \\

\hline
        \ \ \ \  &
        \ \ \ \ $\langle S3, getLocations(), S4, \{\}, true, \{\}, \{\} \rangle$ \\

        \ \ \ \ $\langle S9, matchGPS(), S10, \{(SPS,UMS)\}, true, \{\}, \{\} \rangle$  & \ \ \ \ $\langle S4, \varepsilon, S13, \{\}, true, \{\}, \{\} \rangle$ \\

        \ \ \ \   & \ \ \ \ $\langle S13, \varepsilon, S11, \{(SPS,NMF)\}, true, \{\}, \{\} \rangle$ \\
        \ \ \ \   & \ \ \ \ $\langle S13, \varepsilon, S14, \{(SPS,NMU)\}, true, \{\}, \{\} \rangle$ \\

\hline

         \cellcolor{gray!25}$\textbf{CM}_{\textbf{SPS,NMU}}$ &
         \cellcolor{gray!25}$\textbf{CM}_{\textbf{SPS,NMF}}$ \\

\hline
        \ \ \ \ $\langle S14, notifyUser(), S15, \{\}, true, \{\}, \{\} \rangle$ & \ \ \ \ $\langle S11, notifyFriend(), S12, \{\}, true, \{\}, \{\} \rangle$ \\

        \ \ \ \ $\langle S15, \varepsilon, S16, \{\}, true, \{[S15,(SPS,NMF)]\},$ & \ \ \ \ $\langle S12, \varepsilon, S16, \{\}, true, \{[S12,(SPS,NMU)]\},$ \\

		\ \ \ \ \ \ \ \ $\{[S12,(SPS,NMF)]\} \rangle$ & \ \ \ \ \ \ \ \ $\{[S15,(SPS,NMU)]\} \rangle$ \\

        \ \ \ \ $\langle S16, \varepsilon, S21, \{(SPS,NMF)\}, true, \{\}, \{\} \rangle$ & \ \ \ \ $\langle S16, \varepsilon, S21, \{\}, true, \{\}, \{\} \rangle$ \\

		\ \ \ \ $\langle S22, startItin(), S23, \{\}, true, \{\}, \{\} \rangle$ & \ \ \ \ $\langle S21, startItin, S22, \{(SPS,NMU)\}, true, \{\}, \{\} \rangle$ \\

        \ \ \ \ $\langle S23, \varepsilon, Final, \{\}, true, \{\}, \{\} \rangle$ &  \\

\hline
\end{tabular}
}
\end{center}
\vspace{-0.5cm}
\caption{Coordination Models Tuples}\label{tab:coordModelsTuples}
\end{table}


\noindent Going back to our illustrative example, Table~\ref{tab:coordModelsTuples} reports the CMs derived from the CeFM in Figure~\ref{fig:cltsSample}. Intuitively, the first tuple in $CM_{IM,UMS}$ specifies that $CD_{IM,UMS}$ can perform the operation $getUserPref$ from the source state $S5$ to the target state $S6$; whereas, the second tuple specifies that $CD_{IM,UMS}$ can step over $S6$ and reach $S7$, from where alternative branches can be undertaken. Then, as specified by the third and fourth tuple, $CD_{IM,UMS}$ can reach either $S20$ or $S9$ according to the evaluation of the related conditions, i.e., $\neg shareEnabled$ or $shareEnabled$, respectively. This means that, after $getUserPref$ has been requested by $IM$ and forwarded to $UMS$, in the case $shareEnabled$ holds, $CD_{IM,UMS}$ uses the fourth tuple $\langle S7, \varepsilon, S9, \{(IM,SPS)\}, shareEnabled, \{\}, \{\} \rangle$ to step over the alternative state $S7$, reach $S9$, and inform $CD_{IM,SPS}$ about the new global state $S9$.


Considering the second tuple $\langle$$S15$,$\epsilon$,$S16$,$\{\}$,$true$,$\{[S15,(SPS,NMF)]\}$,$\{[S12,(SPS,NMF)]\}$$\rangle$
in $CM_{SPS,NMU}$, $CD_{SPS,NMU}$ notifies $S15$ to $CD_{SPS,NMF}$ and waits for receiving the notification by $CD_{SPS,NMF}$ about $S12$.
On the other hand, considering the tuple
$\langle$$S12$,$\epsilon$,$S16$,$\{\}$,$true$,$\{[S12,(SPS,NMU)]\}$,$\{[S15,(SPS,NMU)]\}$$\rangle$ in $CM_{SPS,NMF}$, $CD_{SPS,NMF}$ notifies $S12$ to $CD_{SPS,NMU}$ and waits for receiving the notification by $CD_{SPS,NMU}$ about $S15$. Note that the same considerations apply in case different parallel threads of the same CD are involved in a join state.

\vspace{-0.4cm}

%% file: enforcement.tex
\section{Choreography enforcement via distributed coordination}
\label{sec:enforcement}

\vspace{-0.3cm}

In this section we formalize the distributed coordination algorithm that describes the coordination logic that each CD has to perform by relying on its CM. 
The algorithm inherits the distributed mutual exclusion principle and leverages some foundational notions (such as happened-before relation, partial ordering, and time-stamps) of the algorithm in~\cite{lamport:1978}. More precisely, the enforcement of mutual exclusive access to a single resource in~\cite{lamport:1978} is scaled up to the enforcement of concurrent task flows according to arbitrarily complex choreographies.
In the style of~\cite{lamport:1978}, the most appropriate way to present the algorithm is to define rules that each delegate $CD_{i,j}$ follows in a distributed setting, when its supervised service $S_i$ sends a request to perform a task $t$ with $S_j$ (without relying on any central synchronizing entity or shared memory). The actions defined by each rule are assumed to form a single event (i.e., each rule is atomic). As detailed below, a CD is a {\em reactive} entity that, at each iteration of the algorithm, WAITs for receiving either a request from the service it supervise or NOTIFY/UPDATE messages from the other CDs.

The rules locally characterize the collaborative behavior of the CDs at run-time from a clear {\it one-to-many} point of view. The most important aspect here is that, upon reaching a join state, the \emph{involved} CDs notify and wait for each other in order to synchronize their execution.
At run time, NOTIFY messages (together with a simple notion of {\em priority} $P_{i,j}$, initially associated to each $CD_{i,j}$) are used to realize join states as distributed synchronization points. Then, when a join state has been realized, and hence all the parallel executions have been synchronized, the CD with highest priority is in charge of notifying the CDs that, according to the choreography, are allowed to proceed.
%
%
Each $CD_{i,j}$ sends NOTIFY messages according to the coordination information contained by the sixth element of the $CM_{i,j}$ tuples. The ``co-related'' WAIT primitive is instead performed according to the information contained in the seventh element of the tuples. In our implementation, the correct co-relation among WAIT primitives and NOTIFY (or UPDATE) messages is realized by means of dedicated queues that, for each WAIT, buffer the expected NOTIFYs (or UPDATEs). $CD_{i,j}$ uses UPDATE messages whenever the current global state of $C$ changes according to the performed coordination. By considering the coordination information represented by the fourth element of its tuples, $CD_{i,j}$ sends an UPDATE message in order to inform, about the state change, those CDs whose execution can progress from the new current global state. Thus, if a $CD_{h,k}$ receives a request to perform a task $t$ while its local execution is in a state where $t$ is not allowed, then it waits for receiving an UPDATE message on a state from which $t$ is allowed.


\smallskip
\smallskip
In Table~\ref{tab:algo}, we defined three rules. In brief, Rule 1 governs the exchange of business-level messages by considering the cases in which: 
a CD receives a message that is allowed by $C$, hence forwarding it to the interested service as soon as the CD reaches the enabling state of $C$; and a CD steps over $\epsilon$-tasks  by exploiting the procedure defined in Table~\ref{tab:step-over}. Leveraging the distribution step of Definition~\ref{cmgen_def} (which is performed statically), the procedure is capable to handle in a uniform way all kinds of states, i.e., plain, fork, join, alternative, and loop states. At coordination-level, Rule 2 and Rule 3 regulate the exchange of the UPDATE and NOTIFY messages, respectively. Specifically, Rule 2 updates the local current state of the CD according to the current global state of $C$ as received through the UPDATE message. Rule 3 allows the CD to proceed only after all the needed NOTIFY messages have been received. For the sake of presentation, the algorithm assumes that each alternative branch has at least one (and only one) enabling condition that evaluates to true. Furthermore, each participant does not request to perform a task that is not present in the choreography specification. However, our implementation of the algorithm is able to detect these cases raising specific exceptions.

\smallskip

\noindent Adhering to Definition~\ref{cmgen_def}, the algorithm takes as input $CM_{i,j}$. Thus, let $\tau$ be a tuple in $CM_{i,j}$:


\begin{itemize}
\item $\tau[{\tt src}]$ (resp., $\tau[{\tt trg}]$) is the enabling source (resp., target) state of the transition labeled with $\tau[{\tt allowedOp}]$;

\item $\tau[{\tt allowedOp}]$ is the operation that can be performed by $S_i$ when in the source state;

\item $\tau[{\tt allowedService}]$ is the set of CDs (and hence, of supervised services)
that can proceed from the target state;

\item $\tau[{\tt cond}]$ is the PL predicate associated to an alternative path, which allows a CD to proceed on that path whenever it holds;

\item $\tau[{\tt toBeNotified}]$ (resp., $\tau[{\tt toBeWaited}]$) is the set of CDs that, progressing on parallel paths together with $CD_{i,j}$, must be notified (resp., waited for) as soon as $CD_{i,j}$ reaches the state joining the paths. 
\end{itemize}

\noindent At the beginning, all CDs are waiting for receiving an initiating UPDATE message to internally set the state(s) from which they can perform an operation. These messages are sent by an activating component, specifically developed within CHOReOS, called Enactment Engine (\url{https://github.com/choreos/enactment_engine}). The latter is also in charge of automatically deploying the CDs.



\begin{small}
\begin{longtable}{|p{0.9\textwidth}|}

\hline
\vspace{0,05cm}

\underline{\bf Rule 1:} Upon receiving, in the current state $s$ of $C$, a request from $S_i$ to perform a task $t$ with $S_j$, \\


    \vspace{-0,5cm}
    \begin{itemize} \item[] {\bf 1.1} {\bf if} there exists $\tau\in CM_{i,j}$ {\it s.t.} $\tau[{\tt src}]=s$ and $\tau[{\tt allowedOp}]=t$ (i.e., $t$ is allowed from $s$) {\bf then} \end{itemize} \\

        \vspace{-1,0cm}
        \begin{itemize} \item[] \begin{itemize} \item[] {\bf 1.1.1} $CD_{i,j}$ forwards to $S_j$ the message initiating $t$, and gives the control back to $S_i$~\footnote{Note that (\emph{i}) the Service-CD 
        interaction is synchronous; 
        (\emph{ii}) the CD-CD interaction is either synchronous or asynchronous; 
        (\emph{iii}) the CD-CD 
        exchange of coordination information is asynchronous.}; 
        \end{itemize} \end{itemize} \\

        \vspace{-1,0cm}
        \begin{itemize} \item[] \begin{itemize} \item[] {\bf 1.1.2} $CD_{i,j}$ updates $s$ to $\tau[{\tt trg}]$; \end{itemize} \end{itemize} \\

        \vspace{-1,0cm}
        \begin{itemize} \item[] \begin{itemize} \item[] {\bf 1.1.3} {\bf for all} $(h,k)$ $\in$ $\tau[{\tt allowedService}]$, $CD_{i,j}$ sends UPDATE($\tau[{\tt trg}]$) to $CD_{h,k}$; \end{itemize} \end{itemize} \\

        \vspace{-1,0cm}
        \begin{itemize} \item[] \begin{itemize} \item[] {\bf 1.1.4} {\bf \texttt{StepOver}}($s$); \end{itemize} \end{itemize} \\

    \vspace{-1,0cm}
    \begin{itemize} \item[] {\bf else if} $\tau[{\tt src}]\neq s$ for each $\tau\in CM_{i,j}$ {\it s.t.} $\tau[{\tt allowedOp}]=t$ (i.e., $t$ is not allowed from $s$) {\bf then} \end{itemize}
    \vspace{-0,5cm}

    \begin{itemize} \item[] \begin{itemize} \item[] {\bf 1.1.5} $CD_{i,j}$ WAITs for receiving UPDATE($\tau[{\tt src}]$), hence temporarily blocking $S_i$~\footnote{See Rule 2. In other words, $CD_{i,j}$ is in a source state, say $s^\prime$, different from $s$ and it is waiting for being notified for $s^\prime$ to become the new current state of $C$. Note that, since the service-to-CD interaction is synchronous, the task $t$ is pending and $S_i$ is waiting for receiving the control back from $CD_{i,j}$.} \end{itemize} \end{itemize} \\

\hline
\vspace{0,05cm}

\underline{\bf Rule 2:} When $CD_{i,j}$ receives  UPDATE($state$) for some $state$ that it was waiting for \textbf{then} \\

\vspace{-0,5cm}
\begin{itemize} \item[] {\bf 2.1} $CD_{i,j}$ updates $s$ to $state$; \end{itemize} \\

\vspace{-1,0cm}
\begin{itemize} \item[] {\bf 2.2} {\bf if} a task $t$ is pending~\footnote{That is, upon receiving a request to perform $t$, $CD_{i,j}$ was waiting to receive UPDATE($state$).} \textbf{then} \textbf{goto} Rule 1.1; \end{itemize} \\

\hline
\vspace{0,05cm}

\underline{\bf Rule 3:} When $CD_{i,j}$ receives  all NOTIFY($predecessor$, $CD_{h,k}$, $join$) it was waiting for\footnote{See Rule 1.2.4.} {\bf then}

%

\vspace{-0,3cm}
\begin{itemize} \item[] {\bf 3.1} {\bf if} $CD_{i,j}$ has the highest priority (i.e., for all $CD_{h,k}$, $P_{i,j}$$>$$P_{h,k}$) \textbf{then} \end{itemize} \\

    \vspace{-1,0cm}
    \begin{itemize} \item[] \begin{itemize} \item[] {\bf 3.1.1} $CD_{i,j}$ updates $s$ to $join$; \end{itemize} \end{itemize} \\

    \vspace{-1,0cm}
    \begin{itemize} \item[] \begin{itemize} \item[] {\bf 3.1.2} let $\tau\in CM_{i,j}$ be {\it s.t.} $\tau[{\tt src}]=join$; \end{itemize} \end{itemize} \\

    \vspace{-1,0cm}
    \begin{itemize} \item[] \begin{itemize} \item[] {\bf 3.1.3} {\bf for all} $(h,k)$ $\in$ $\tau[{\tt allowedService}]$, $CD_{i,j}$ sends UPDATE($\tau[{\tt trg}]$) to $CD_{h,k}$; \end{itemize} \end{itemize} \\

\vspace{-1,0cm}
\begin{itemize} \item[] {\bf 3.2} {\bf if} a task $t$ is pending \textbf{then} \textbf{goto} Rule 1.1; \end{itemize} \\

\hline
\caption{Rule-based description of the algorithm} \label{tab:algo}
\vspace{-0,2cm}
\end{longtable}
\end{small}

\vspace{-0.5cm}

\begin{small}
\begin{longtable}{|p{0.9\textwidth}|}

\hline
\vspace{0,05cm}

{\bf \texttt{StepOver}}($s$): \\

    \vspace{-0,6cm}
    \begin{itemize} \item[] {\bf while} there exists $\tau\in CM_{i,j}$ {\it s.t.} $\tau[{\tt src}]=s$ and $\tau[{\tt allowedOp}]=\tau.\varepsilon.\tau$ (i.e., it is possible to proceed from $s$ only by an internal task) and there exists $\tau^\prime\in CM_{i,j}$ {\it s.t.} $\tau^\prime[{\tt src}]=\tau[{\tt trg}]$ and $\tau^\prime[{\tt cond}]$ holds {\bf do} \end{itemize} \\
		
		\vspace{-1,0cm}
		\begin{itemize} \item[] \begin{itemize} \item[] {\bf \texttt{StepOver}}($\tau^\prime[{\tt trg}]$); \end{itemize} \end{itemize} \\
		
        \vspace{-1,0cm}
        \begin{itemize} \item[] \begin{itemize} \item[] {\bf for all} $[s,(h,k)]$ $\in$ $\tau[{\tt toBeNotified}]$, $CD_{i,j}$ sends NOTIFY($s$, $CD_{h,k}$, $\tau[{\tt trg}]$) to $CD_{h,k}$; \end{itemize} \end{itemize} \\

        \vspace{-1,0cm}
        \begin{itemize} \item[] \begin{itemize} \item[] {\bf for all} $[s^{\prime\prime},(h,k)]$ $\in$ $\tau[{\tt toBeWaited}]$, $CD_{i,j}$ WAITs for receiving NOTIFY($s^{\prime\prime}$, $CD_{h,k}$, $\tau[{\tt trg}]$) from $CD_{h,k}$, hence temporarily blocking $S_i$ (see Rule 3); \end{itemize} \end{itemize} \\


        \vspace{-1,0cm}
        \begin{itemize} \item[] \begin{itemize} \item[] {\bf for all} $(h,k)$ $\in$ $\tau^\prime[{\tt allowedService}]$, $CD_{i,j}$ sends UPDATE($\tau^\prime[{\tt trg}]$) to $CD_{h,k}$; \end{itemize} \end{itemize}
        \vspace{-1,0cm}
\\


\hline
\caption{{\bf \texttt{StepOver}} procedure} \label{tab:step-over}
\end{longtable}
\end{small}

\begin{figure}[h]
  \center
    \includegraphics[width=0.8\textwidth]{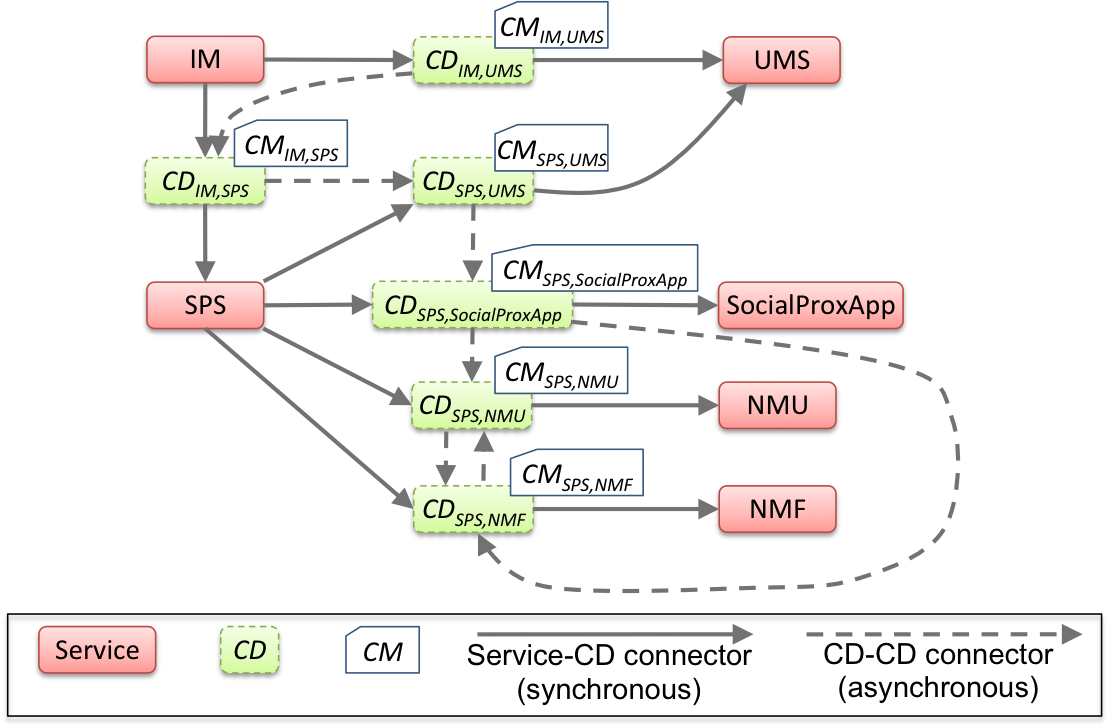}
\caption{Architecture of the enforced Distributed Social Proximity Network scenario}
  \label{fig:example-arch}
\end{figure}

\noindent Going back to our illustrative example, Figure~\ref{fig:example-arch} shows the software architecture of the system that realizes the choreography specified by the CeFM shown in Figure~\ref{fig:cltsSample}. This system is composed of the services that instantiate the choreography participant roles (e.g., {\tt IM}, {\tt SPS}), the automatically synthesized CDs (e.g., $CD_{IM,SPS}$, $CD_{SPS,UMS}$) together with their CMs (e.g., $CM_{IM,SPS}$, $CM_{SPS,UMS}$), synchronous connectors enabling Service-CD interaction by exchanging business level messages (e.g., the {\em getUserPref} and {\em notifyUser} operations), and asynchronous connectors enabling CD-CD interaction by exchanging additional coordination messages (e.g., NOTIFY and UPDATE messages). The deployment information needed to realize the structure of the depicted architecture is automatically synthesized by exploiting the set of generated CMs.

As introduced above, at the beginning, $CD_{IM,SPS}$ is waiting for receiving an UPDATE on state $S9$ (of the CeFM); $CD_{IM,UMS}$ is waiting for receiving an UPDATE on $S5$; $CD_{SPS,UMS}$ is waiting for receiving an UPDATE on $S10$; $CD_{SPS,SocialProxApp}$ is waiting for receiving an UPDATE on $S3$; $CD_{SPS,NMU}$ is waiting for receiving an UPDATE on either $S14$ or $S22$; and $CD_{SPS,NMF}$ is waiting for receiving an UPDATE on either $S11$ or $S21$. Again, this information is automatically synthesized out of the generated CMs.
Once the synthesized CDs are deployed, by obeying the constraints of the architecture shown in Figure~\ref{fig:example-arch}, the CHOReOS Enactment Engine sends UPDATE($S5$) to $CD_{IM,UMS}$, hence starting the execution of the choreography.

\begin{landscape}
\begin{figure}[h]
  \center
    \includegraphics[height=0.9\textheight, width=1.3\textwidth]{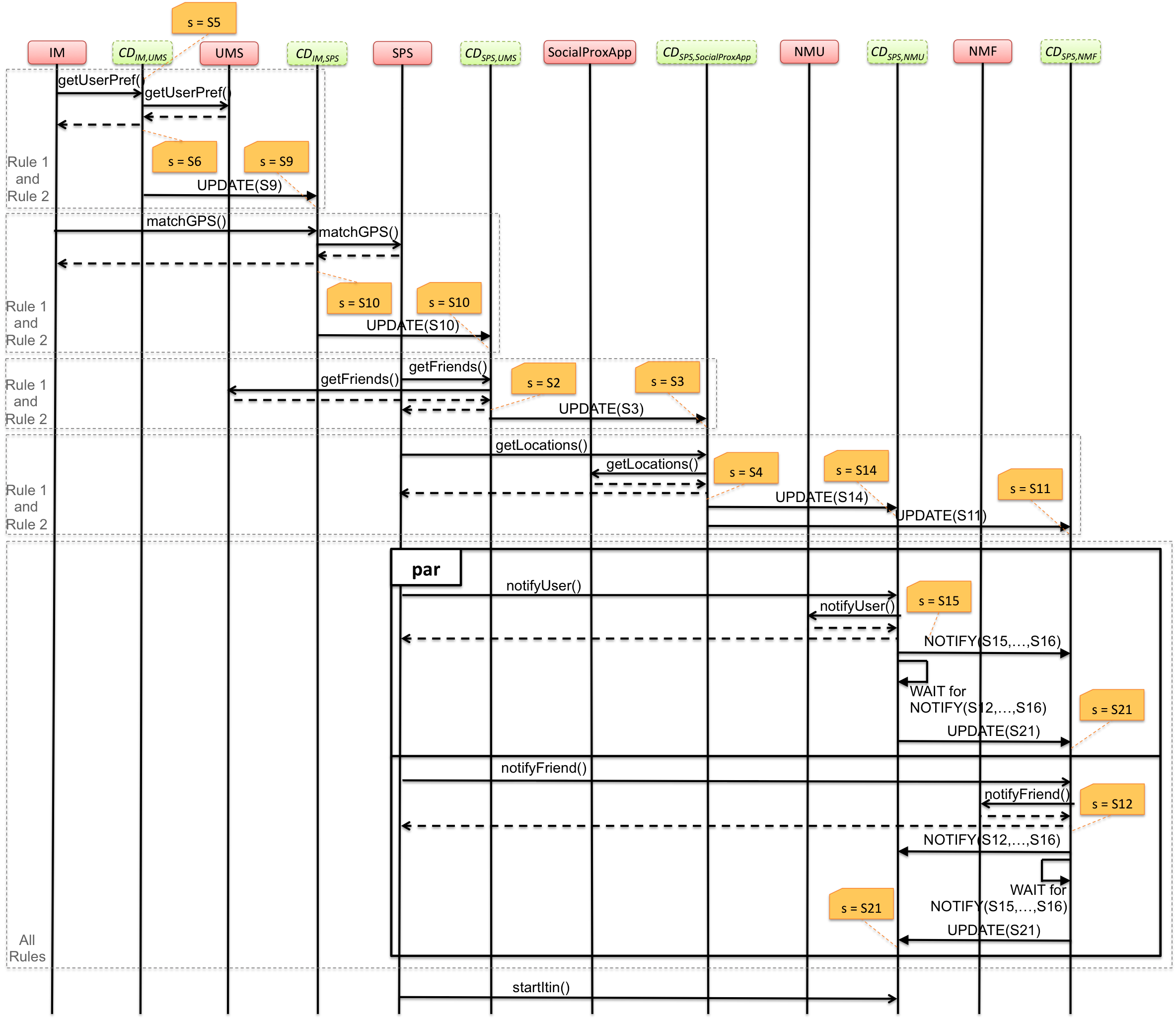}
\caption{Messages exchange for the Distributed Social Proximity Network scenario}
  \label{fig:sequence}
\end{figure}
\end{landscape}

To illustrate the execution of the distributed coordination algorithm at work on our example, Figure~\ref{fig:sequence} shows a sequence diagram representing an excerpt\footnote{From the initial state to S21, and by assuming that both {\em shareEnabled} and {\em friendFound} hold.} of the exchange of business- and coordination-level messages among the participants and their supervising CDs. As it is evident from the diagram, the collaboration of the synthesized CDs coordinates the interaction among the participant services in order to let them perform only the interactions specified by the CeFM, hence preventing, e.g., the undesired interaction mentioned in Section~\ref{sec:scenario}. We recall that this undesired interaction concerns the case in which a notified friend takes on the created itinerary when the other friend has not been notified yet.

%% file: correctness.tex
\section{Correctness} \label{sec:correctness}

\vspace{-0.3cm}

In this section, 
we give a rigorous characterization of undesired interactions, and prove that our enforcement method prevents them.


\smallskip
\smallskip

\noindent For our distributed coordination algorithm to be correct, it should never happen that:

\begin{description}
\item[\textbf{(i)}] a CD accepts a request to perform a task that is not allowed in the current state;

\item[\textbf{(ii)}] although the task is allowed, the current state (a) has been reached by an alternative state whose condition does not hold; (b) is a loop entry state but the guard of the loop does not hold; (c) is a loop exit state but the guard still holds; 
(d) is the successor of a join state but some parallel flows must still be completed; (e) leads to a predecessor of a join state where, after performing the requested task, each involved CD is waiting for the others.
\end{description}


\smallskip
\smallskip

\noindent These considerations lead to the following definition that characterizes the way undesired interactions can be detected at run-time. That is, an undesired interaction occurs whenever an {\em undesired operation}, as defined by Definition~\ref{def:undesired_operation}, occurs. In the definition, we make use of the following notations. 
$CM_{i,j}(s)$ denotes the set of tuples in $CM_{i,j}$ with $s$ as first element. $CM_{i,j}(s)$$\upharpoonright_{Op}$ denotes the set of choreography operations (different from $\varepsilon$) appearing as second element of tuples in $CM_{i,j}(s)$ (i.e., all the operations allowed from $s$). $CM_{i,j}(s)$$\upharpoonright_{Wait}$ denotes the union of valid triples in $Wait_{siblings(s)}$ (i.e., all the predecessors of $s$ that must be waited for at run-time). The predicate $\sim$$Notify_{i,j}$ holds if and only if $CD_{i,j}$ has not received yet all NOTIFY messages that it is waiting for. 

\smallskip
\smallskip

\begin{definition}[Undesired operation] \label{def:undesired_operation} Let $CM_{i,j}$ be the CM generated for the delegate $CD_{i,j}$ out of the CeFM. Let $s$$\in$$S_C$ be the current state reached by $CD_{i,j}$ during its execution. A choreography operation $\alpha$$=$$r_i.t.r_j$ (for some $r_i,r_j$$\in$$R_C$ and $t$$\in$$\mathcal{T}$) is an {\em undesired operation} in $s$ if and only if one of the following conditions hold:
\begin{itemize}
\item {\bf undesired task:} $\alpha$$\notin$$CM_{i,j}(s)$$\upharpoonright_{Op}$;
\item {\bf undesired flow:} $\alpha$$\in$$CM_{i,j}(s)$$\upharpoonright_{Op}$ $\wedge$ one of the following conditions hold:
\begin{itemize}
\item {\bf invalid alternative:} $\exists$$s^{alt}$$\in$$S^{ALT}_C$,$\rho$$\in$$\Phi$ $:$ $s^{alt}$$\freccia{\rho}_C$$s$ $\wedge$ $\neg$$\rho$;
\item {\bf invalid loop:} $\exists$$s^{loop}$$\in$$S^{LOOP}_C$,$\rho$$\in$$\Phi$ $:$ $s^{loop}$$\freccia{\rho}_C$$s$ $\wedge$ $\neg$$\rho$;
\item {\bf missed loop:} $\exists$$s^{loop}$$\in$$S^{LOOP}_C$,$\rho$$\in$$\Phi$,$s^\prime$ $:$ $s^{loop}$$\freccia{}_C$$s$ $\wedge$ $s^{loop}$$\freccia{\rho}_C$$s^\prime$ $\wedge$ $\rho$;
\vspace{-0.1cm}
\item {\bf missed join:} $\exists$$s^{join}$$\in$$S^{JOIN}_C$ $:$ $s^{join}$$\freccia{}_C$$s$ $\wedge$ $\sim$$Notify_{i,j}$;
\item {\bf deadlocking join:} $\exists$$s^{join}$$\in$$S^{JOIN}_C$,$s^\prime$ $:$ $s$$\freccia{\alpha}_C$$s^\prime$$\freccia{}$$s^{join}$ $\wedge$ $CD_{i,j}$ is in $s^\prime$ $\wedge$ $\sim$$Notify_{i,j}$ $\wedge$ $\forall$$(s^{\prime\prime},h,k)$$\in$$CM_{i,j}(s)$$\upharpoonright_{Wait}$ $:$ $CD_{h,k}$ is in $s^{\prime\prime}$ $\wedge$ $\sim$$Notify_{h,k}$.
\end{itemize}
\end{itemize}
\end{definition}

\noindent \textbf{Correctness proof (by contradiction)} -- Let us suppose that, at run-time, the collaboration of the synthesized CDs and the participants performs an undesired operation. By Definition~\ref{def:undesired_operation}, this means that either an undesired task or an undesired flow has been performed. By referring to the distributed coordination algorithm defined in Section~\ref{sec:enforcement}, the {\bf undesired task} condition contradicts {\bf Rule 1.1}. Now, let us consider all the cases that characterize the {\bf undesired flow} condition. The {\bf invalid alternative}, {\bf invalid loop}, {\bf missed loop}, and {\bf missed join} conditions contradict the execution of procedure {\bf \texttt{StepOver}}; finally, the {\bf deadlock} condition contradicts the combined execution of {\bf \texttt{StepOver}} with {\bf Rule 3} and {\bf Rule 2}. In any case, we deduce a contradiction and, hence, the correctness proof is given meaning that the interaction among the generated CDs and the participants prevents
any kind of undesired interaction.

%% file: related.tex
\section{Related work} \label{sec:related}


The method formalized in this paper is related to a large number of other approaches that have been considered in the literature in the domains of {\em service-oriented and component-based engineering}, and {\em controller synthesis and priority scheduling}. We discuss here only the ones that are closer to the automated choreography enforcement problem.

\smallskip
\smallskip

\noindent {\it \textbf{Service-oriented and component-based engineering}} -- There are many approaches aiming at composing services by means of BPEL, WSCI, or WS-CDL choreographers~\cite{BP06,CGL08,MPT08,MPM08,PBL08,Sal08,SBF07}. The common idea underlying these approaches is to assume a high-level specification of the requirements that the choreography has to fulfil and a behavioural specification of the participants. From these two assumptions, by applying data and control-flow analysis, the BPEL, WSCI or WS-CDL description of a centralized choreographer is automatically derived so that it satisfies the specified requirements. In particular, in~\cite{SBF07}, the authors propose an approach to derive service implementations from a choreography specification.
In~\cite{Sal08}, the authors assume that some services are reused and propose an approach to exploit wrappers to make the reused services match the choreography. Most of the previous approaches concern orchestration, which is a form of composition different form choreography. The former focuses on centralized coordination, hence resulting in a monolithic composition. Instead, the latter is a means for composing services in a fully distributed way.

Despite the fact that the works described in~\cite{Sal08,SBF07,Bultan:2011,Bultan:2010,Basu-Bultan-POPL:12} focus on choreography, they consider the problem of checking choreography realizability. Note that it is a fundamentally different problem from the one considered in this paper. In fact, our approach is reuse-oriented and aims at restricting, by means of the synthesized CDs, the interaction behaviour of the discovered (third-party) services in order to realize the specified choreography. Differently, the approaches described in~\cite{Sal08,SBF07,Bultan:2011,Bultan:2010,Basu-Bultan-POPL:12} are focused on verifying whether the set of services, required to realize a given choreography, can be easily implemented by simply considering the role-based local views of the specified choreography. That is, this verification does not aim at synthesizing the coordination logic, which is needed whenever the collaboration among the participants leads to global interactions that violate the choreography behaviour.

Similarly, later work in~\cite{Gwen:12,pascal12} valuably advances the state-of-the-art results. Then, most closely to our work, in~\cite{GwenPascalFASE:13} the authors presents a framework for verifying choreographies using model- and equivalence-checking techniques. The framework enables the verification of some analysis tasks, i.e., repairability, realizability, conformance, synchronizability, and control for enforcing realizability. In order to check in sequence the system synchronizability and realizability using equivalence checking, distributed controllers are generated through an iterative process presented in~\cite{GwenGudemann:2012}.
Counterexamples are exploited to refine the behaviour of the controllers by adding new synchronization messages until both synchronizability and realizability can be enforced.

Works in the area of the synthesis of runtime monitors from automata are described in~\cite{monitoring:2004,Simmonds09}. Note that runtime monitoring is mostly focused on the detection of undesired behaviours, while runtime enforcement focuses on their prevention.

In the context of {\em Reo connectors}, a number of related works are worth to be discussed. 
The works described in~\cite{fmco08,tsc13,isola08} discuss different approaches to extract coordination/choreography specifications from BPMN diagrams, UML state diagrams, and UML activity diagrams. 
Automated synthesis of coordination/choreography specifications from scenario-based specification is addressed in~\cite{scp11,coordination05}. The works described in~\cite{pdp14,esocc13,esocc12} are more specific with respect to the problem of distributed enforcement/implementation of choreographies. In particular, in~\cite{pdp14}, the authors provide a systematic and rigorous way of constructing hybrid (i.e., partially-distributed) connector implementations for distributed orchestrations. 
In~\cite{esocc13}, the authors define a new product operator for {\em Constraint Automata} (CA) whose computation at run-time requires only relatively simple distributed algorithms. 
The work in~\cite{esocc12} describes an hybrid approach for the automatic code generation of orchestrations with Reo. 
The main focus of these works is on hybrid enforcement approaches. 

\smallskip
\smallskip

\noindent {\it \textbf{Controller synthesis and priority scheduling}} -- The controller synthesis problem is conceptually similar to ours (yet technically different). That is, given a system specification and a property, synthesize a controller that, once put in parallel with the system, enforces the satisfaction of the property. In~\cite{Baier:2011a}, the authors describe a compositional approach for efficiently constructing a centralized controller. Then, in~\cite{Baier:2011b}, they describe how to decompose a monolithic controller into a network of Reo connectors. The former focuses on compositionality of the synthesis rather than on controller distribution, whereas the latter focuses on data-flow coordination.

In~\cite{Graf:2012a}, the synthesis of a distributed controller is made decidable by model-checking {\em knowledge} properties and exploiting {\em temporary synchronization}. Similar concepts have been exploited also for monitoring global properties in distributed systems~\cite{Graf:2011b}. Model-checking knowledge properties for addressing scheduling problems in distributed systems has been exploited also in~\cite{Peled:2009a}. The work described in~\cite{Peled:2011a} shares the same idea of using additional interactions in order to synthesize \enquote{joint} knowledge of several concurrent and distributed entities. Our notions of CM and coordination information resemble their notions of knowledge and temporary synchronization, respectively.

Priorities define precedence relations between component actions. Since priorities can be used to restrict the behaviour of a system in order to avoid undesired states, our work is related to priority scheduling. In~\cite{Bensalem:2011a}, a priority synthesis algorithm is presented. It aims at synthesizing a system that is deadlock-free or satisfies some safety property. In~\cite{Bensalem:2011b}, the authors describe an algorithm to synthesize local sets of priorities hence addressing distribution of the control logic. Priority-based synthesis of distributed controllers has been addressed also in~\cite{Graf:2011a} and, in the Web service orchestration setting, in~\cite{Graf:2009a}. These works have been either conceived in the context of the {\em BIP framework}, which is quite different from ours, or adopting its main concepts and operations.

%% file: conclusions.tex
\section{Conclusions and future work} \label{sec:conclusions}


We have formalized a method for the distributed enforcement of service choreographies. It takes as input a choreography specification given in the form of an automata-based model, called CeFM, derived from a BPMN2 choreography diagram. This specification is automatically decomposed into a set of CMs that contain the information, local to each participant service, needed to coordinate the global interaction of all the participants from outside so to realize the specified choreography. Relying on the generated CMs, a distributed coordination algorithm has been defined with the aim of characterizing the coordination logic that has to be performed by additional software entities, called CDs. Once interposed among the participant services, CDs collaborate with each other to enforce the specified choreography.

We have illustrated the applicability of our method by means of a distributed social proximity network scenario. We have proven correctness of the defined algorithm with respect to preventing those interactions that do not belong to the choreography specification. To this end, we have rigorously characterized the notion of undesired interaction.

The formalized method is implemented as part of a model-based tool chain released under the FISSi initiative (\url{http://www.ow2.org/view/Future_Internet}) to support the development of choreography-based systems in the CHOReOS EU project. The proposed method has been also applied to other real-world use cases in the {\em Airport}, and {\em Marketing and Sale} domains. Refer to the CHOReO\emph{Synt} web site \url{http://choreos.disim.univaq.it} for documentation.
The application to these use cases has shown that the method is viable and the overhead due to the exchange of synchronization information is negligible. Indeed, in the worst case, it is bound by the number of tasks times the number of participants.

Beyond the modelling of message-based interaction, BPMN2 Choreography Diagrams provide constructs for specifying events. As future work, we plan to extend our definitions of CeFM, and related CMs, to account for event-based coordination. As a consequence, the distributed coordination algorithm formalized in Section~\ref{sec:enforcement} must also be revised. As mentioned in Section~\ref{sec:introduction}, our distributed coordination algorithm is general in the sense that it is service-independent. As a further future work, we will leverage our preliminary results in~\cite{serene:2013,icse:2013} to release the assumption that the participants already match the corresponding role specifications. This means that we have to define an extended version of CDs that are able to address both coordination and adaptation issues in a modular manner, hence still promoting separation of concerns. Moreover, we plan to prove other notions of correctness beyond undesired interaction prevention, e.g., progress or permissiveness, and deadlock freedom beyond deadlocking joins.
Last but not least, not all choreographies can be enforced by our method and, as further future work, we should formally characterize the conditions under which a choreography is enforceable by our method.

\section*{Acknowledgments} 

This work is supported by the Ministry of Education, Universities and Research, prot. 2012E47TM2 (project IDEAS - Integrated Design and Evolution of Adaptive Systems), and was supported by the European Community's Seventh Framework Programme FP7/2007-2013 under grant agreement number 257178 (project CHOReOS - Large Scale Choreographies for the Future Internet - \url{www.choreos.eu}).

We acknowledge the anonymous reviewers for their thoughtful feedbacks and valuable suggestions that allowed us to significantly improve the paper.